\documentclass[12pt,titlepage]{article}
\usepackage[USenglish]{babel}
\usepackage{sectsty}
\usepackage{amsfonts}
\usepackage{amsmath}
\usepackage{amssymb}
\usepackage{geometry}
\usepackage{apacite}
\usepackage{setspace}
\usepackage{apalike}
\usepackage{float}
\usepackage{portland}

\input{tcilatex}
\begin{document}

\begin{center}
\textbf{A} \textbf{Bayesian Nonparametric IRT Model\footnote{%
This material is based upon work supported by National Science Foundation
grant SES-1156372 from the Program in Methodology, Measurement, and
Statistics. The first author gives thanks to Wim J. van der Linden and Brian
Junker for feedback on this work.}}

George Karabatsos

February 11, 2015

\bigskip
\end{center}

\textbf{\noindent Abstract:}\ \ This paper introduces a flexible Bayesian
nonparametric Item Response Theory (IRT) model, which applies to dichotomous
or polytomous item responses, and which can apply to either unidimensional
or multidimensional scaling. This is an infinite-mixture IRT model, with
person ability and item difficulty parameters, and with a random intercept
parameter that is assigned a mixing distribution, with mixing weights a
probit function of other person and item parameters. As a result of its
flexibility, the Bayesian nonparametric IRT model can provide outlier-robust
estimation of the person ability parameters and the item difficulty
parameters in the posterior distribution. The estimation of the posterior
distribution of the model is undertaken by standard Markov chain Monte Carlo
(MCMC) methods based on slice sampling. This mixture IRT model is
illustrated through the analysis of real data obtained from a teacher
preparation questionnaire, consisting of polytomous items, and consisting of
other covariates that describe the examinees (teachers). For these data, the
model obtains zero outliers and an R-squared of one. The paper concludes
with a short discussion of how to apply the IRT model for the analysis of
item response data, using menu-driven software that was developed by the
author.

\section{Introduction\label{Introduction}}

Given a set of data, consisting of person'$\,$s individual responses to
items of a test, an item response theory (IRT)\ model aims to infer each
person's ability on the test, and to infer the test item parameters. In
typical applications of an IRT\ model, each item response is categorized
into one of two or more categories. For example, each item response may be
scored as either correct (1)\ or incorrect (0). From this perspective, a
categorical regression model, which includes person$\,$ ability parameters
and item difficulty parameters, provides an interpretable approach to
inferring from item response data. One basic example is the Rasch (1960%
\nocite{Rasch60}) model. This model can be characterized as a logistic
regression model, having the dichotomous item score as the dependent
variable.\ The predictors (covariates)\ of this model include $N$\ person$\,$
indicator (0,1) variables, corresponding to regression coefficients that
define the person$\,$ ability parameters; and include $I$ item indicator
(0,-1) variables, corresponding to coefficients that define the item
difficulty parameters.

In many item response data sets, there are observable and unobservable
covariates that influence the item responses, in addition to the person$\,$
and item factors. If the additional covariates are not fully accounted for
in the given IRT\ model, then the estimates of person$\,$ ability and item
difficulty parameters can become noticeably biased. Such biases can be (at
least)\ partially-alleviated by including the other, observable covariates
into the IRT\ (regression)\ model, as control variables. However, for most
data collection protocols, it is not possible to collect data on all the
covariates that help determine the item responses (e.g., due to time,
financial, or ethical constraints). Then, the unobserved covariates, which
influence the item responses, can bias the estimates of the ability and item
parameters in an IRT\ model that does not account for these covariates.

A flexible mixture IRT\ model can provide robust estimates of person$\,$
ability parameters and item difficulty parameters, by accounting for any
additional unobserved latent covariates\ that influence the item responses.
Modeling flexibility can be maximized through the use of a Bayesian
nonparametric (BNP)\ modeling approach.

In this chapter we present a BNP\ approach to infinite-mixture IRT modeling,
based on the general BNP\ regression model introduced by Karabatsos and
Walker (2012\nocite{KarabatsosWalker12c}). We then illustrate this model,
called the BNP-IRT\ model, through the analysis of real item response data.
The analysis was conducted using a menu-driven (point-and-click) software,
developed by the author (Karabatsos 2014a, 2014b\nocite{Karabatsos14a}\nocite%
{Karabatsos14b}).

In the next section, we give a brief overview of the concepts of mixture IRT
modeling, and BNP infinite-mixture modeling. Then in Section 3, we introduce
our basic, BNP-IRT\ model. This is a regression model consisting of person$%
\, $ ability and item difficulty parameters, constructed via the appropriate
specification of person$\,$ and item indicator predictor variables, as
mentioned above. While the basic model assumes dichotomous item scores and
unidimensional person$\,$ ability, our model can be easily extended to
handle polytomous responses (with item response categories not necessarily
ordered), extra person$\,$-level and/or item-level covariates, and/or
multidimensional person$\,$ ability parameters. In Section 4, we describe
the Markov chain Monte Carlo (MCMC)\ methods that can be used to estimate
the posterior distribution of the model parameters. (This is a highly
technical section which can be skipped when reading this chapter). In
Section 5, we describe methods for evaluating the fit of our BNP-IRT\ model.
Section 6 provides an empirical illustration of the BNP-IRT model through
the analysis of polytomous response data. The data were obtained from an
administration of a questionnaire that was designed to measure teacher
preparation. Section 7 ends with a brief overview of how to use the
menu-driven software to perform data analysis using the BNP-IRT\ model. That
section also includes a brief discussion of how to extend the BNP-IRT\ model
for cognitive IRT.

The remained of this chapter makes use of the following notational
conventions. Let $\boldsymbol{U}=(U_{1},\ldots ,U_{i},\ldots
,U_{I})^{\intercal }$ denote a random vector for the scores on a test with $%
I $ items. A realized value of the item response vector is denoted by $%
\boldsymbol{u}=(u_{1},\ldots ,u_{i},\ldots ,u_{I})^{\intercal }$.\ We assume
that each item $i=1,\ldots ,I$ has $m_{i}+1$ possible discrete-valued
scores, indexed by $u=0,1,\ldots ,m_{i}$.

We use lower cases to denote a probability mass function (pmf) of a value $u$
discrete random variable (or vector, $\mathbf{u}$) or a probability density
function (pdf) of a value $u$ of a continuous random variable (or $\mathbf{u}
$), such as $f(u)$ or $f(\mathbf{u})$, respectively. The given pmf (or pdf)\ 
$f(u)$ corresponds to a cumulative distribution function (cdf), denoted by
upper case $F(u)$, which gives the probability that the random variable $U$
does not exceed $u$. $F(u)$ is sometimes more simply referred to as the
distribution function. Thus, for example, $\mathrm{N}$$(\mu ,\sigma ^{2})$, 
\textrm{U}$(0,b)$, \textrm{IG}$(a,b)$ and \textrm{Be}$(a,b)$ (or cdfs $%
\mathrm{N}$$(\cdot \,|\,\mu ,\sigma ^{2})$, \textrm{U}$(\cdot \,|\,0,b)$, 
\textrm{IG}$(\cdot \,|\,a,b)$ and \textrm{Be}$(\cdot \,|\,a,b)$,
respectively), denote the univariate normal, uniform, inverse-gamma, and
beta distribution functions, respectively. They correspond to pdfs $\mathrm{n%
}(\cdot \,|\,\mu ,\sigma ^{2})$, \textrm{u}$(\cdot \,|\,0,b)$, \textrm{ig}$%
(\cdot \,|\,a,b)$, \textrm{be}$(\cdot \,|\,a,b)$, with mean and variance
parameters$\,(\mu ,\sigma ^{2})$, minimum and maximum parameters $(0,b)$,
shape and rate parameters $(a,b)$, and shape parameters $(a,b)$,
respectively. Also, if $\boldsymbol{\beta }$ is a realized value of a $K$%
-dimensional random vector, then $\mathrm{N}(\boldsymbol{\beta }\,|\,\mathbf{%
0},\mathbf{V})$ denotes the cdf of the multivariate ($K$-variate)\ normal
distribution with mean vector of zeros $\mathbf{0}$ and $K\times K$
variance-covariance matrix $\mathbf{V}$, distribution function $\mathrm{n}(%
\mathbf{0},\mathbf{V})$, and corresponding to pdf $\mathrm{n}(\boldsymbol{%
\beta }\,|\,\mathbf{0},\mathbf{V})$. The pmf or pdf\ of $u$ given values of
one or more variables $\mathbf{x}$ is written as $f(u\,|\,\mathbf{x})$ (with
corresponding cdf $F(u\,|\,\mathbf{x})$); given a vector of parameter values 
$\boldsymbol{\zeta }$ is written as $f(u\,|\,\boldsymbol{\zeta })$ (with
corresponding cdf $F(u\,|\,\boldsymbol{\zeta })$), and conditionally on
variables and given parameters is written as $f(u\,|\,\mathbf{x};\boldsymbol{%
\zeta })$ (with corresponding cdf $F(u\,|\,\mathbf{x};\boldsymbol{\zeta })$%
). Also, $\sim $ means "distributed as", $\sim _{ind}$ means "independently
distributed," and $\sim _{iid}$ means "independently and identically
distributed." For example, $U\sim $ $F$, $U\sim _{iid}F(u)$, $U\sim $ $%
F(u\,|\,\mathbf{x};\boldsymbol{\zeta })$, $U\sim $ $F(u\,|\,\boldsymbol{%
\zeta })$, $U\sim _{iid}F(\boldsymbol{\zeta })$, $\boldsymbol{\beta }$ $\sim 
\mathrm{N}(\mathbf{0},\mathbf{V})$, or $\sigma ^{2}\sim $ \textrm{IG}$(a,b)$%
. The preceding notation may replace $U$ by $\mathbf{U}$, replace $F$ by $f$%
, \ replace $\mathrm{N}$ by $\mathrm{n}$, and/or replace \textrm{IG} by 
\textrm{ig.}

\section{Mixture IRT and Bayesian Nonparametrics\label{Bayesian
Nonparametrics and Mixture IRT}}

For any given vector of item response data $\boldsymbol{u}=(u_{1},\ldots
,u_{i},\ldots ,u_{I})^{\intercal }$, a discrete-mixture IRT model admits the
general form%
\begin{equation}
f_{G_{\mathbf{x}}}(\boldsymbol{u}\,|\,\mathbf{x})=\dint f(\boldsymbol{u}\,|\,%
\mathbf{x};\boldsymbol{\beta },\mathbf{\Psi }(\mathbf{x}))\text{\textrm{d}}%
G_{\mathbf{x}}(\mathbf{\Psi })=\dsum\limits_{j=1}^{J}\,f(\boldsymbol{u}\,|\,%
\mathbf{x};\boldsymbol{\beta },\mathbf{\Psi }_{j}(\mathbf{x}))\omega _{j}(%
\mathbf{x}).  \label{RanDensx}
\end{equation}%
conditionally on any given value of a vector of any\ covariates $\mathbf{x}$%
. In this expression, $f(\boldsymbol{u}\,|\,\mathbf{x};\boldsymbol{\beta },%
\mathbf{\Psi }(\mathbf{x}))$ is the kernel of the mixture, and $G_{\mathbf{x}%
}$ is a mixture distribution that may (or may not)\ depend on the same
covariates.

Also, as show in (\ref{RanDensx}), this pmf is based on a mixture of $J$
pmfs $f(\boldsymbol{u}\,|\,\mathbf{x};\boldsymbol{\beta },\mathbf{\Psi }_{j}(%
\mathbf{x}))$, $j=1,\ldots ,J$. Here, $\boldsymbol{\beta }$ is a vector of
(any available)\ fixed parameters that are not subject to the mixture, the $%
\mathbf{\Psi }_{j}(\mathbf{x}),$ $j=1,\ldots ,J$,\ are random parameters
that are subject to the mixture that may be covariate dependent, and $J$ is
the number of mixture components. In addition, $\omega _{j}(\mathbf{x}),$ $%
j=1,\ldots ,J$,\ are mixture weights that sum to one for every given
covariate value $\mathbf{x}\in \mathcal{X}$. The mixture model (\ref%
{RanDensx}) is called a discrete (continuous)\ mixture model if $G_{\mathbf{x%
}}$ is discrete (continuous); it is called a finite (infinite)\ mixture
model if $J$ is finite (infinite).

A simple example is given by the finite mixture Rasch model for dichotomous
item scores (Rost, 1990\nocite{Rost90}, 1991\nocite{Rost91}; von Davier \&
Rost, vol. 1, chap. 23), which assumes that%
\begin{equation}
f(\boldsymbol{u}\,|\,\mathbf{x};\boldsymbol{\beta },\mathbf{\Psi }_{j}(%
\mathbf{x}))=\dprod\limits_{i=1}^{I}\frac{\exp (\theta _{j}-\beta
_{ij})^{u_{i}}}{1+\exp (\theta _{j}-\beta _{ij})},  \label{Finite Rasch mix}
\end{equation}%
with a finite number of $J$ components and mixture weights that are not
covariate-dependent (i.e., $\omega _{j}(\mathbf{x})=\omega _{j}$, $%
j=1,\ldots ,J<\infty $). The ordinary Rasch (1960\nocite{Rasch60}) model for
dichotomous item scores is the special case of the model defined by (\ref%
{RanDensx}) and (\ref{Finite Rasch mix}) for $J=1$.

An infinite-mixture model is given by (\ref{RanDensx}) for $J=\infty $. A
general BNP\ infinite-mixture IRT\ model assumes that the mixture
distribution has the general form%
\begin{equation}
G_{\mathbf{x}}(\cdot )=\dsum\limits_{j=1}^{\infty }\omega _{j}(\mathbf{x}%
)\delta _{\mathbf{\Psi }_{j}(\mathbf{x})}(\cdot ),  \label{GenSeth}
\end{equation}%
where $\delta _{\mathbf{\Psi }}(\cdot )$ denotes a degenerate distribution
with support $\mathbf{\Psi }$.\ This Bayesian model is completed by the
specification of a prior distribution on $\{\mathbf{\Psi }_{j}(\mathbf{x}%
)\}_{j=1,2,\ldots }$, $\{\omega _{j}(\mathbf{x})\}_{j=1,2,\ldots },$ and $%
\boldsymbol{\beta }$ with large supports.

A common example is a Dirichlet process mixed IRT model, which assumes that
the mixing distribution is not covariate-dependent (i.e., $G_{\mathbf{x}%
}(\cdot )=G(\cdot )$), along with a random mixing distribution $G(\cdot )$\
constructed as $G(\cdot )=\tsum\nolimits_{j=1}^{\infty }\omega _{j}\delta _{%
\mathbf{\Psi }_{j}}(\cdot )$ where $\omega _{j}=\upsilon
_{j}\tprod\nolimits_{k=1}^{j-1}(1-\upsilon _{k})$ for random draws $\upsilon
_{j}\sim _{iid}\mathrm{Be}(1,\alpha )$ and $\mathbf{\Psi }_{j}\sim
_{iid}G_{0},$ for $j=1,2,\ldots $. Here, $G$ is a Dirichlet process (DP),
denoted $G\sim \mathrm{DP}(\alpha ,G_{0})$, with baseline parameter $G_{0}$
and precision parameter $\alpha $ (Sethuraman, 1994\nocite{Sethuraman94}).\
The $\mathrm{DP}(\alpha ,G_{0})$\ has mean (expectation)\ $\mathbb{E}%
[G(\cdot )]=G_{0}(\cdot )$ and variance $\mathbb{V}[G(\cdot )]=G_{0}(\cdot
)\{1-G_{0}(\cdot )\}/(\alpha +1)$ (Ferguson, 1973\nocite{Ferguson73}).

An important generalization of the DP\ prior includes the Pitman-Yor
(Poisson-Dirichlet) prior (Ishwaran \&\ James, 2001\nocite{IshwaranJames01}%
), which assumes that $\upsilon _{j}\sim _{iid}$\textrm{Be}$(\alpha
_{1j},\alpha _{2j})$, for $j=1,2,\ldots $, for some $\alpha _{1j}=1-\alpha
_{1}$ and $\alpha _{2j}=\alpha _{2}+j\alpha _{1}$ with $0\leq \alpha _{1}<1$
and $\alpha _{2}>-\alpha _{1}$. The special case defined by $\alpha _{1}=0$
and $\alpha _{2}=\alpha $ results in the \textrm{DP}$(\alpha G_{0})$.

Another important generalization of the DP is given by the Dependent
Dirichlet process (DDP)\ (MacEachern, 1999\nocite{MacEachern99}, 2000\nocite%
{MacEachern00}, 2001\nocite{MacEachern01}), which provides a model for the
covariate-dependent random distribution, denoted $G_{\mathbf{x}}$. The DDP\
model assumes that $G_{\mathbf{x}}\sim $ \textrm{DP}$(\alpha _{\mathbf{x}%
},G_{0\mathbf{x}})$, marginally for each $\mathbf{x}$. Specifically, the
DDP\ defines a covariate-dependent random distribution $G_{\mathbf{x}}$ of
the form given in equation (\ref{GenSeth}), and incorporates this dependence
either through covariate-dependent atoms $\mathbf{\Psi }_{j}(\mathbf{x})$, a
covariate-dependent baseline $G_{0\mathbf{x}}$, and/or covariate-dependent
stick-breaking weights of the form $\omega _{j}(\mathbf{x})=\upsilon _{j}(%
\mathbf{x})\tprod\nolimits_{k=1}^{j-1}(1-\upsilon _{k}(\mathbf{x}))$, for $%
j=1,2,\ldots $. For example, the ANOVA-linear DDP\ (De Iorio et al. 2004%
\nocite{DeIorioMullerRosnerMacEachern04}), denoted $G_{\mathbf{x}}\sim $%
\textrm{\ ANOVA-DDP}$(\alpha ,G_{0},\mathbf{x})$, constructs a dependent
random distribution $G_{\mathbf{x}}(\cdot )=\tsum\nolimits_{j=1}^{\infty
}\omega _{j}\delta _{\mathbf{x}^{\intercal }\boldsymbol{\beta }_{j}}(\cdot )$%
, via covariate-dependent atoms $\mathbf{x}^{\intercal }\boldsymbol{\beta }%
_{j}$, along with $\boldsymbol{\beta }\sim G$ and $G\sim $ \textrm{DP}$%
(\alpha ,G_{0}\mathbf{(}\boldsymbol{\beta }\mathbf{)})$.

Many examples of DP-mixture and DDP mixture IRT models can be found in the
literature (Qin, 1998\nocite{Qin98}; Duncan \&\ MacEachern, 2008\nocite%
{DuncanMacEachern08}; Miyazaki \& Hoshino, 2009\nocite{MiyazakiHoshino09};
Farina et al. 2009\nocite{FarinaQuintanaSanMartinJara09}; San Martin et al.
2011\nocite{SanMartinJaraRolinMouchart11}; San Martin, et al., 2011\nocite%
{SanMartinRolinCastro11}; Karabatsos \&\ Walker, 2012\nocite%
{KarabatsosWalker12a}).

\section{Presentation of the Model\label{Presentation of the Model}}

The BNP-IRT model is a special case of the Bayesian nonparametric regression
(infinite-mixture)\ model introduced by Karabatsos and Walker (2012\nocite%
{KarabatsosWalker12c}). These authors demonstrated that the model tended to
have better predictive performance relative to DP-mixed and DDP\ mixed
regression models. As will be shown, the BNP-IRT model is suitable for
dichotomous or polytomous item responses.

First, we present the basic BNP-IRT\ model for dichotomous item responses.
Let $\mathcal{D}=\{(u_{pi},\mathbf{x}_{pi})_{i=1}^{I}\}_{p=1}^{P}$ denote a
set of item-response data, including dichotomous responses $u_{pi}\in
\{0,1\} $. Also, $\mathbf{x}_{pi}$ denotes a covariate vector that describes
person$\,$ $p=1,...,P$ and item $i=1,\ldots ,I$.

The basic BNP-IRT\ model is defined as 
\begin{subequations}
\label{BNP-IRT}
\begin{eqnarray}
f(\mathcal{D}\,|\,\mathbf{X};\,\boldsymbol{\zeta })
&=&\dprod\limits_{p=1}^{P}\dprod\limits_{i=1}^{I}f(u_{pi}\,|\,\mathbf{x}%
_{pi};\,\boldsymbol{\zeta })  \label{Like1} \\
f(u_{pi}\,|\,\mathbf{x}_{pi};\,\boldsymbol{\zeta }) &=&P(U_{pi}=1\,|\,%
\mathbf{x}_{pi};\,\boldsymbol{\zeta })^{u_{pi}}[1-P(U_{pi}=1\,|\,\mathbf{x}%
_{pi};\,\boldsymbol{\zeta })]^{1-u_{pi}}  \label{Like2} \\
\Pr (U &=&1\,|\,\mathbf{x};\,\boldsymbol{\zeta })=1-F^{\ast }(0\,|\,\mathbf{x%
};\,\boldsymbol{\zeta })=\dint\limits_{0}^{\infty }f(u_{pi}^{\ast }\,|\,%
\mathbf{x}_{pi};\,\boldsymbol{\zeta })\text{\textrm{d}}u^{\ast }
\label{Like3} \\
&=&\dint\limits_{0}^{\infty }\tsum\limits_{j=-\infty }^{\infty }\text{%
\textrm{n}}(u^{\ast }\,|\,\mu _{j}+\mathbf{x}_{pi}^{\mathbf{\intercal }}%
\boldsymbol{\beta },\sigma ^{2})\omega _{j}(\mathbf{x}_{pi}\mathbf{;}%
\boldsymbol{\beta }_{\omega },\sigma _{\omega })\text{\textrm{d}}u^{\ast }
\label{Like4} \\
\omega _{j}(\mathbf{x;}\boldsymbol{\beta }_{\omega },\sigma _{\omega })
&=&\Phi \left( \dfrac{j-\mathbf{x}^{\mathbf{\intercal }}\boldsymbol{\beta }%
_{\omega }}{\sigma _{\omega }}\right) -\Phi \left( \dfrac{j-1-\mathbf{x}^{%
\mathbf{\intercal }}\boldsymbol{\beta }_{\omega }}{\sigma _{\omega }}\right)
\label{mixweights} \\
(\mu _{j},\sigma _{\mu }^{2}) &\sim &\text{$\mathrm{N}$}(\mu
_{j}\,|\,0,\sigma _{\mu }^{2})\text{\textrm{U}}(\sigma _{\mu
}\,|\,0,b_{\sigma \mu })  \label{priors1} \\
(\boldsymbol{\beta },\boldsymbol{\beta }_{\omega }) &\sim &\mathrm{N}(%
\boldsymbol{\beta }\,|\,\mathbf{0},\sigma ^{2}v\mathrm{diag}(\infty ,\mathbf{%
J}_{NI}^{\intercal }))\text{$\mathrm{N}$}(\boldsymbol{\beta }_{\omega }\,|\,%
\mathbf{0},\sigma _{\omega }^{2}v_{\omega }\mathbf{I}_{NI+1})
\label{priors2} \\
(\sigma ^{2},\sigma _{\omega }^{2}) &\sim &\text{\textrm{IG}}(\sigma
^{2}\,|\,a_{0}/2,a_{0}/2)\text{\textrm{IG}}(\sigma _{\omega
}^{2}\,|\,a_{\omega }/2,a_{\omega }/2).  \label{priors3}
\end{eqnarray}%
Under the model, the data likelihood is given by equations (\ref{Like1})-(%
\ref{mixweights}) given parameters $\boldsymbol{\zeta }=(\boldsymbol{\mu }%
,\sigma _{\mu },\boldsymbol{\beta },\boldsymbol{\beta }_{\omega },\sigma
^{2},\sigma _{\omega })$ with $\boldsymbol{\mu }=(\mu _{j})_{j=-\infty
}^{\infty }$. By default, the model assumes that $\mathbf{x}_{pi}$ is a
binary indicator vector with $NI+1$ rows, having constant (1)\ in the first
entry, a \textquotedblleft 1\textquotedblright\ in entry $p+1$ to indicate
person $p$, and \textquotedblleft $-1$\textquotedblright\ in entry $i+(p+1)$
to indicate item $i$. Specifically, each vector $\mathbf{x}_{pi}$ is defined
by 
\end{subequations}
\begin{equation*}
\mathbf{x}_{pi}=(1,\mathbf{1}(p=1),\ldots ,\mathbf{1}(p=N),-\mathbf{1}%
(i=1),\ldots ,-\mathbf{1}(i=I))^{\intercal },
\end{equation*}%
where $\mathbf{1}(\cdot )$ denotes the indicator\ (0,1) function. Then, in
terms of the coefficient vector $\boldsymbol{\beta }=(\beta _{0},\beta
_{1},\ldots ,\beta _{PI})$, each coefficient $\beta _{p+1}=\theta _{p}$
represents the ability of person $p=1,\ldots ,P$. Likewise, each coefficient 
$\beta _{i+p+1}$ represents the difficulty of item $i=1,\ldots ,I$. The
covariate-dependent mixture weights $\omega _{j}(\mathbf{x})$ in (\ref%
{mixweights}) are specified by a cumulative ordered probits regression,
based on the choice of a standard normal cdf for $\Phi \left( \cdot \right) $
with latent mean $\mathbf{x}^{\mathbf{\intercal }}\boldsymbol{\beta }%
_{\omega }$ and variance $\sigma _{\omega }^{2}$, for the "ordinal
categories" $j=0,\pm 1,\pm 2,\ldots $, where coefficient vector $\boldsymbol{%
\beta }_{\omega }$ contains additional person$\,$ parameters and item
parameters.

As shown in (\ref{priors1})--(\ref{priors3}), the Bayesian model parameters $%
\boldsymbol{\zeta }$ have joint prior density 
\begin{subequations}
\label{Prior Density}
\begin{eqnarray}
\pi (\boldsymbol{\zeta }) &=&\tprod\limits_{j=-\infty }^{\infty }\mathrm{n}%
(\mu _{j}\,|\,0,\sigma _{\mu }^{2})\text{\textrm{u}}(\sigma _{\mu
}\,|\,0,b_{\sigma \mu })\mathrm{n}(\boldsymbol{\beta }\,|\,\mathbf{0},\sigma
^{2}\mathrm{diag}(\infty ,v\mathbf{J}_{NI}^{\intercal })) \\
&&\times \mathrm{n}(\boldsymbol{\beta }_{\omega }\,|\,\mathbf{0},\sigma
_{\omega }^{2}v_{\omega }\mathbf{I}_{NI+1})\text{\textrm{ig}}(\sigma
^{2}\,|\,a_{0}/2,a_{0}/2)\text{\textrm{ig}}(\sigma _{\omega
}^{2}\,|\,a_{\omega }/2,a_{\omega }/2),
\end{eqnarray}%
where $\mathbf{J}_{NI}^{\intercal }$ denotes the vector of $NI$ ones, and $%
\mathbf{I}_{NI+1}$ is the identity matrix of dimension $NI+1$. As shown in (%
\ref{Prior Density}), the full specification of their prior density\ relies
on the choice of the parameters $(b_{\sigma \mu },v,a_{0},v_{w},a_{w})$. In
Section 6, where we illustrate the BNP-IRT\ model through the analysis of a
real item response data set, we suggest some useful default choices for
these prior parameters.

As shown by the model equations in (\ref{Like1})-(\ref{Like4}), the item
response function $\Pr (U=1\,|\,\mathbf{x};\,\boldsymbol{\zeta })$ is
modeled by a covariate($\mathbf{x}$)-dependent location mixture of normal
distributions for the latent variables $u_{pi}^{\ast }$. The random
locations $\mu _{j}$ of this mixture corresponds to mixture weights $\omega
_{j}(\mathbf{x})$, $j=0,\pm 1,\pm 2,\ldots $. Conditionally on a covariate
vector, $\mathbf{x}_{pi}$ and model parameters, the latent mean and variance
of the mixture can be written as: 
\end{subequations}
\begin{eqnarray*}
\mathbb{E}[U_{pi}^{\ast }\,|\,\mathbf{x}_{pi};\boldsymbol{\beta },%
\boldsymbol{\beta }_{\omega },\sigma ^{2},\sigma _{\omega }] &=&\mu
_{pi}^{\ast }=\dsum\limits_{j=-\infty }^{\infty }(\mu _{j}+\mathbf{x}_{pi}^{%
\mathbf{\intercal }}\boldsymbol{\beta })\omega _{j}(\mathbf{x}_{pi}\mathbf{;}%
\boldsymbol{\beta }_{\omega },\sigma _{\omega }), \\
\mathbb{V}[U_{pi}^{\ast }\,|\,\mathbf{x}_{pi};\boldsymbol{\beta },%
\boldsymbol{\beta }_{\omega },\sigma ^{2},\sigma _{\omega }]
&=&\dsum\limits_{j=-\infty }^{\infty }\{[(\mu _{j}+\mathbf{x}_{pi}^{\mathbf{%
\intercal }}\boldsymbol{\beta })-\mu _{pi}^{\ast }]^{2}+\sigma ^{2}\}\omega
_{j}(\mathbf{x}_{pi}\mathbf{;}\boldsymbol{\beta }_{\omega },\sigma _{\omega
}),
\end{eqnarray*}%
respectively (Marron \& Wand, 1992\nocite{MarronWand92}).

The BNP-IRT\ model can be viewed as an extension of the DP-mixed binary
logistic generalized linear model (Mukhopadhyay \&\ Gelfand, 1997\nocite%
{MukhopadhyayGelfand97}).\ In terms of the responses $u$, the extension can
be written as%
\begin{eqnarray*}
f(u\,|\,\mathbf{x}) &=&\dsum\limits_{j=1}^{\infty }\dfrac{\exp (\mu _{j}\,+\,%
\mathbf{x}^{\mathbf{\intercal }}\boldsymbol{\beta })^{u}}{1+\exp (\mu
_{j}\,+\,\mathbf{x}^{\mathbf{\intercal }}\boldsymbol{\beta })}\omega _{j} \\
\omega _{j} &=&\upsilon _{j}\tprod\nolimits_{k=1}^{j-1}(1-\upsilon _{k}) \\
\upsilon _{j} &\sim &\mathrm{Be}(1,\alpha ),\text{ }j=1,2,\ldots \\
\mu _{j} &\sim &\mathrm{N}(0,\sigma _{\mu }^{2}),\text{ }j=1,2,\ldots \\
\boldsymbol{\beta } &\sim &\mathrm{N}(\mathbf{0},\mathbf{\Sigma }_{%
\boldsymbol{\beta }}).
\end{eqnarray*}%
This model thus defines a mixture of logistic cdfs for the inverse link
function, with weights $\omega _{j}$ that are not covariate-dependent. In
contrast, as shown in (\ref{Like3})--(\ref{Like4}), the BNP-IRT\ model in (%
\ref{BNP-IRT}) is based on a mixture of normal cdfs for the inverse link
function. The BNP-IRT\ model is more flexible than the DP\ model, because
the former uses covariate-dependent mixture weights, as shown in (\ref%
{mixweights}).

In other words, if $\mu _{j}=0$ for all $j$, then the BNP-IRT\ model reduces
to the Rasch IRT model with "normal-ogive"\ response functions; all items
are assumed to have common slope (discrimination) parameter that is
proportional to $1/\sigma $. Nonzero values of $\mu _{j}$, along with the
covariate-dependent mixture weights $\omega _{j}(\mathbf{x;}\boldsymbol{%
\beta }_{\omega },\sigma _{\omega })$, for $j=0,\pm 1,\pm 2,\ldots $, allows
the BNP-IRT\ model to shift the location of each response function across
persons and items. Value of $\mu _{j}>0$ ($\mu _{j}<0$) shifts the response
function to the left (right). The BNP-IRT\ model allows for this shifting in
a flexible manner, accounting for any outlying responses (relative to a
normal-ogive Rasch model). This feature enables inferences of person$\,$and
item parameters from the BNP-IRT\ that model are robust against such
outliers.

According to Bayes' theorem, a set of data $\mathcal{D}$ updates of the
prior probability density $\pi (\boldsymbol{\zeta })$ in (\ref{Prior Density}%
) leads to posterior probability density 
\begin{equation*}
\pi (\boldsymbol{\zeta }\,|\,\mathcal{D})=\frac{f(\mathcal{D}\,|\,\mathbf{X}%
;\,\boldsymbol{\zeta })\pi (\boldsymbol{\zeta })}{\dint f(\mathcal{D}\,|\,%
\mathbf{X};\,\boldsymbol{\zeta })\pi (\boldsymbol{\zeta })\mathrm{d}%
\boldsymbol{\zeta }}.
\end{equation*}%
Also, conditionally on $(\mathbf{x}_{pi},\mathcal{D})$, the posterior
predictive pmf and the posterior expectation ($\mathbb{E}$) and variance ($%
\mathbb{V}$) of the item response $U_{pi}$ are given by%
\begin{eqnarray}
f(u_{pi}\,|\,\mathbf{x}_{pi},\mathcal{D}) &=&\dint f(u_{pi}\,|\,\mathbf{x}%
_{pi};\,\boldsymbol{\zeta })\pi (\boldsymbol{\zeta }\,|\,\mathcal{D})\mathrm{%
d}\boldsymbol{\zeta },  \label{PPD1} \\
\mathbb{E}[U_{pi}\,|\,\mathbf{x}_{pi},\mathcal{D}] &=&f(U_{pi}=1\,|\,\mathbf{%
x}_{pi},\mathcal{D})=f(1\,|\,\mathbf{x}_{pi},\mathcal{D}),  \label{PPD2} \\
\mathbb{V}[U_{pi}\,|\,\mathbf{x}_{pi},\mathcal{D}] &=&f(1\,|\,\mathbf{x}%
_{pi},\mathcal{D})[1-f(1\,|\,\mathbf{x}_{pi},\mathcal{D})],  \label{PPD3}
\end{eqnarray}%
respectively.

It is straightforward to extend the BNP-IRT\ regression model to other types
of response data by making appropriate choices of covariate vector $\mathbf{x%
}$ (corresponding to coefficients $\boldsymbol{\beta },\boldsymbol{\beta }%
_{\omega }$). Such extensions are described as follows:

\begin{enumerate}
\item Suppose that for each item $i=1,\ldots ,I$ the responses are each
scored in more than two categories, say $m_{i}+1$ nominal or ordinal
categories denoted as $u^{\prime }=0,1,\ldots ,m_{i}$, with $u^{\prime }=0$
the reference category. Then the model can be extended to handle such
polytomous item responses using the Begg and\ Gray (1984\nocite{BeggGray1984}%
) method. Specifically, the model would assume the response to be defined by 
$u_{pi}=\mathbf{1}(u_{pi}^{\prime }>0)$ each covariate vector $\mathbf{x}%
_{pi}$ to be defined by a binary indicator vector:%
\begin{eqnarray*}
\mathbf{x}_{pi} &=&(1,\mathbf{1}(p=1),\ldots ,\mathbf{1}(p=N),\mathbf{1}(i=1)%
\mathbf{1}(u_{pi}^{\prime }=1),\ldots ,\mathbf{1}(i=I)\mathbf{1}%
(u_{pi}^{\prime }=1),\ldots , \\
\mathbf{1}(i &=&1)\mathbf{1}(u_{pi}^{\prime }=m_{i}),\ldots ,\mathbf{1}(i=I)%
\mathbf{1}(u_{pi}^{\prime }=m_{i}))^{\intercal }.
\end{eqnarray*}%
Then in terms of coefficient vector $\boldsymbol{\beta }=(\beta _{0},\beta
_{1},\ldots ,\beta _{1+p+m^{\ast }I})$, coefficient $\beta _{1+p}=\theta
_{p} $, $p=1,\ldots ,P,$ represents the latent ability of person $p$ and the
coefficient $\beta _{1+p+(u-1)I+i}$ represents the latent difficulty of item 
$i=1,\ldots ,I$ and category $u=1,\ldots ,m^{\ast }$, where\ $m^{\ast
}=\max_{i}m_{i}$.

\item If the data has additional covariates $(x_{1},\ldots ,x_{q})$ which
describe either the persons (e.g., socioeconomic status), test items (e.g.,
item type), or type of response (e.g., response time), associated with each
person $p$ and item $i$, then these covariates can be added as the last $q$
elements to each of the covariate vectors $\mathbf{x}_{pi}$, such that $%
\mathbf{x}_{pi}=(\ldots ,x_{1i},\ldots ,x_{qi})^{\intercal }$, $p=1,\ldots
,P $ and $i=1,\ldots ,I$. Then, specific elements of coefficient vector $%
\boldsymbol{\beta }$, namely the elements $\beta _{k}$, $k=\dim (\boldsymbol{%
\beta })-q+1,\ldots ,\dim (\boldsymbol{\beta })$, would represent the
associations of the $q$ covariates with the responses.

\item Similarly, suppose that given test consists of measuring one or more
of $D\leq I$ measurement dimensions. Then we can extend the model to
represent such multidimensional items, by including $D$ binary (0,1)
covariates into the covariate vectors $\mathbf{x}_{pi},$ $p=1,\ldots ,P$ and 
$i=1,\ldots ,I$, such that the first set of elements of $\mathbf{x}_{pi}$
defined by%
\begin{eqnarray*}
\mathbf{x}_{pi} &=&(1,\mathbf{1}(p=1)\mathbf{1}(d_{i}=1),\ldots ,\mathbf{1}%
(p=N)\mathbf{1}(d_{i}=1),\ldots ,\mathbf{1}(p=1)\mathbf{1}(d_{i}=D),\ldots ,
\\
\mathbf{1}(p &=&N)\mathbf{1}(d_{i}=D),\ldots )^{\intercal },
\end{eqnarray*}%
where $d_{i}\in \{1,\ldots ,D\},$ denotes the measurement dimension of item $%
i$. Then specific elements of the coefficient vector $\boldsymbol{\beta }$,
namely the elements $\beta _{k}$, for $k=2,\ldots ,ND+1$, indicate each
person's ability on dimension $d=1,\ldots ,D$.
\end{enumerate}

\section{Parameter Estimation}

By using latent-variable Gibbs sampling methods for Bayesian
infinite-mixture models (Kalli et al. 2011\nocite{KalliGriffinWalker11}), it
is possible to conduct exact MCMC sampling from the posterior distribution
of the BNP-IRT model parameters. More specifically, introducing latent
variables $(\underline{u}_{pi},z_{pi}\in \mathbb{Z},u_{pi}^{\ast }\in 
\mathbb{R})_{N\times I}$ and a fixed decreasing function such as $\xi
_{l}=\exp (-l)$, the conditional likelihood of the BNP-IRT\ model can be
written as%
\begin{equation}
\dprod\limits_{p=1}^{P}\dprod\limits_{i=1}^{I}\mathbf{1}(0<\underline{u}%
_{pi}<\xi _{|z_{pi}|})\xi _{|z_{pi}|}^{-1}\text{\textrm{n}}(u_{pi}^{\ast
}\,|\,\mu _{z_{pi}}+\mathbf{x}{}_{pi}^{\mathbf{\intercal }}\boldsymbol{\beta 
},\sigma ^{2})\omega _{z_{pi}}(\mathbf{x}_{i}^{\mathbf{\intercal }}%
\boldsymbol{\beta }_{\omega },\sigma _{\omega }).  \label{LatLike}
\end{equation}%
For each $(p,i)$, after marginalizing over the latent variables in (\ref%
{LatLike}) we obtain the original model likelihood $f(u_{pi}\,|\,\mathbf{x}%
_{pi};\,\boldsymbol{\zeta })$ in (\ref{Like1}). Importantly, conditionally
on the latent variables, the infinite-dimensional BNP-IRT\ model can be
treated as a finite-dimensional model, which then makes the task of MCMC\
sampling feasible (of course, a even a computer cannot handle an infinite
number of parameters). Given all variables, save the latent variables $%
(z_{i})_{i=1}^{n}$, the choice of each $z_{i}$\ has finite maximum value $%
\pm N_{\max }$, where $N_{\max }=\max_{p}[\max_{i}\{\max_{j}\mathbb{I}(%
\underline{u}_{pi}<\xi _{j})\,|\,j\,|\}]$.

Then standard MCMC\ methods can be used to sample the full conditional
posterior distributions of each latent variable and model parameter
repeatedly for a sufficiently large number of times, $S$. If the prior $\pi (%
\boldsymbol{\zeta })$ is proper (Robert \&\ Casella, 2004, sect. 10.4.3%
\nocite{RobertCasella04}), then, for $S\rightarrow \infty $, this sampling
process constructs a discrete-time Harris ergodic Markov chain%
\begin{equation*}
\{((\underline{u}_{pi}^{(s)}),(z_{pi}^{(s)}),(z_{pi}^{\ast (s)}),\boldsymbol{%
\zeta }^{(s)}=((\underline{u}_{pi}),(z_{pi}),\boldsymbol{\mu },\sigma _{\mu
},\boldsymbol{\beta },\sigma ^{2},\boldsymbol{\beta }_{\omega },\sigma
_{\omega })^{(s)}\}_{s=1}^{S},
\end{equation*}%
which, upon after marginalizing out all the latent variables $(\underline{u}%
_{pi}^{(s)}),(z_{pi}^{(s)}),(z_{pi}^{\ast (s)}),$ has the posterior
distribution $\Pi (\boldsymbol{\zeta }\,|\,\mathcal{D}_{n})$ as its
stationary distribution (for definitions, see Meyn \&\ Tweedie, 1993\nocite%
{MeynTweedie93}; Nummelin, 1984\nocite{Nummelin84}; Roberts \&\ Rosenthal,
2004\nocite{RobertsRosenthal04}). (The next paragraph provides more details
about the latent variables, $z_{pi}^{\ast (s)}$).

The full conditional posterior distribution are as follows: the one of $%
\underline{u}_{pi}$ is \textrm{u}$(\underline{u}_{pi}\,|\,0,\xi _{|z_{pi}|})$%
; $u_{pi}^{\ast }$ has a truncated normal distribution; the one of $z_{pi}$
is a multinomial distribution independently for $p=1,\ldots ,P$ and $%
i=1,\ldots ,I$; the full conditional distribution of $\mu _{j}$ is a normal
distribution (sampled using a Metropolis-Hastings algorithm), independently
for $j=-N_{\max },\ldots ,N_{\max }$; $\sigma _{\mu }$ can be sampled using
a slice sampling algorithm involving a stepping-out procedure (Neal, 2003%
\nocite{Neal03b}); the one $\boldsymbol{\beta }$ is multivariate normal
distribution; and the full conditional posterior distribution of $\sigma
^{2} $ is inverse-gamma. Also, upon sampling of truncated normal latent
variables $z_{pi}^{\ast }$ that have full conditional densities proportional
to $\mathrm{n}(z_{pi}^{\ast }\,|\,\mathbf{x}_{i}^{\mathbf{\intercal }}%
\boldsymbol{\beta }_{\omega },\sigma _{\omega })\mathbf{1}%
(z_{pi}-1<z_{pi}^{\ast }<z_{pi})$, independently for $p=1,\ldots ,P$ and $%
i=1,\ldots ,I$, the full conditional posterior distribution of $\boldsymbol{%
\beta }_{\omega }$ is multivariate normal distribution and the one of $%
\sigma _{\omega }^{2}$ is inverse-gamma distribution. For further details of
the MCMC algorithm, see Karabatsos and Walker (2012\nocite%
{KarabatsosWalker12c}).

In practice, obviously only a MCMC\ chain based on a finite number $S$\ can
be generated. The convergence of finite MCMC\ chains to samples from
posterior distributions can be assessed using the following two procedures
(Geyer, 2011\nocite{Geyer11}):\ (i) viewing univariate trace plots of the
model parameters to evaluate MCMC\ mixing (Robert \&\ Casella, 2004\nocite%
{RobertCasella04}); and (ii)\ conducting a batch-mean (or subsampling)\
analysis of the finite chain, which would provide 95\%\ Monte Carlo
Confidence intervals (MCCIs)\ of all the posterior mean and quantile
estimates of the model parameters (Flegal \& Jones, 2011\nocite%
{FlegalJones11}). Convergence can be confirmed both by trace plots that look
stable and "hairy" and 95\%\ MCCIs that, for all practical purposes, are
sufficiently small. If convergence is not attained for the current choice of 
$S$ samples of a MCMC\ chain, additional MCMC\ samples should be generated
until convergence is obtained.

\section{Model Fit \label{Model Fit Evaluation}}

The fit of the BNP-IRT model to a set of item response data, $\mathcal{D}$,
can be assessed on the basis of its posterior predictive pmf, defined in (%
\ref{PPD1}).

More specifically, the fit to a given response $u_{pi}$ can be assessed by
its standardized response residual%
\begin{equation*}
r_{pi}=\dfrac{u_{pi}-\mathbb{E}[U_{pi}\,|\,\mathbf{x}_{pi},\mathcal{D}]}{%
\sqrt{\mathbb{V}_{n}[U_{pi}\,|\,\mathbf{x}_{pi}]}}.
\end{equation*}%
Response $u_{pi}$ can be judged to be an outlier when $|r_{pi}|\,\ $is
greater than exceeds two or three.

A global measure of the predictive fit of a regression model, indexed by $%
\underline{m}\in \{1,\ldots ,M\}$, is provided by the mean-squared
predictive error criterion%
\begin{equation*}
D(\underline{m})=\dsum\limits_{p=1}^{P}\dsum\limits_{i=1}^{I}\{u_{pi}-%
\mathbb{E}[U_{pi}\,|\,\mathbf{x}_{pi},\mathcal{D}]\}^{2}+\dsum%
\limits_{p=1}^{P}\dsum\limits_{i=1}^{I}\mathbb{V}_{n}[U_{pi}\,|\,\mathbf{x}%
_{pi}\mathbf{,}\underline{m}].
\end{equation*}%
(Laud \&\ Ibrahim, 1995\nocite{LaudIbrahim1995}; Gelfand \&\ Ghosh, 1998%
\nocite{GelfandGhosh98}). The first term of $D(\underline{m})$ measures the
goodness-of-fit ($\mathrm{Gof}(m)$) of the model to the data, while its
second term is a penalty for model complexity. Among a set of $\underline{m}%
=1,\ldots ,M$ that is compared, the model with the highest predictive
accuracy for the data set $\mathcal{D}$ is identified as the one with the
smallest value of $D(\underline{m})$.

The proportion of variance explained by the regression model is given by the 
$R$-squared ($R^{2}$) statistic%
\begin{equation*}
R^{2}=1-\dfrac{\tsum\nolimits_{p=1}^{P}\tsum\nolimits_{i=1}^{I}\{u_{pi}-%
\mathbb{E}[U_{pi}\,|\,\mathbf{x}_{pi},\mathcal{D}]\}^{2}}{%
\tsum\nolimits_{p=1}^{P}\tsum\nolimits_{i=1}^{I}\{u_{pi}-\overline{u}\}^{2}},
\end{equation*}%
where $\overline{u}=\frac{1}{PI}\tsum\nolimits_{p=1}^{P}\tsum%
\nolimits_{i=1}^{I}u_{pi}$.

The standardized residuals $r_{pi}$, the $D(\underline{m})$ criterion, and $%
R^{2}$ can each be estimated as a simple by-product of an MCMC\ algorithm.

\section{Empirical Example\label{Empirical Example}}

Using the BNP-IRT model, we analyzed a set of polytomous response data
obtained from the 2006 \textit{Progress in International Reading Literacy
Study}. A total of $N=244$ fourth-grade U.S. teachers rated their own
teaching preparation level in a ten-item questionnaire ($I=10$). Each item
was scored on a scale ranging from zero to two.

For this questionnaire, the latent person$\,$ ability was assumed to
represent the level of teaching preparation. The ten items addressed the
following areas: education level (named CERTIFICATE), English LANGUAGE,
LITERATURE, teaching reading (PEDAGOGY), PSYCHOLOGY, REMEDIAL reading,
THEORY of reading, children's language development (LANGDEV), special
education (SPED), and second language (SECLANG) learning. The CERTIFICATE
item was scored on a scale of 0 = bachelor's, 1 = master's, 2 = doctoral,
while the other 9 questionnaire items were each scored on a scale consisting
of 0 = not at all, 1 = overview or introduction to topic, and 2 = area of
emphasis. Each of the ten items described a type of training for literacy
teachers, as prescribed by the \textit{National Research Council} (2010%
\nocite{NationalResearchCouncil10}).

We considered three additional covariates for the BNP-IRT model, namely AGE
level (scored in nine ordinal categories), FEMALE\ status, and Miss:FEMALE,
an indicator (0,1)\ of missing value for FEMALE\ status. Overall, $2,419$ of
the total possible $2,440$ item were observed. Three of the $244$ teachers
had missing values for FEMALE, which were imputed using information from the
observed values of all the variables mentioned above.

Given that each of the 10 items item was scored on a polytomous scale\ (3
categories), and that we were interested in additional covariates over and
beyond the person-indicator and item-indicator covariates, we analyzed the
data using the BNP-IRT\ model, using extensions \#1 and \#2 of the basic
BNP-IRT\ model in Section 3 above. Also, the parameters of the prior pdf (%
\ref{Prior Density}) of the model were chosen as $(b_{\sigma \mu
},v,a_{0},v_{w},a_{w})=(1,10,1000,1,.01)$.

To estimate the posterior distribution of the BNP-IRT\ model parameters, we
ran the MCMC\ sampling algorithm in Section 4 for $62,000$ iterations. We
used $12,000$ MCMC samples for posterior inference, retaining every fifth
sample beyond the first $2,000$ iterations (burn-in) to obtain (pseudo-)
independence between them. Trace plots for the univariate parameters
displayed adequate mixing (i.e., exploration of the posterior distribution),
and a batch-mean (subsampling) analysis of the $12,000$ MCMC samples
revealed 95\%\ Monte Carlo Confidence intervals of the posterior mean and
quantile estimates (reported below) that typically had half-widths less than 
$.2$. If desired, smaller half-widths could have been obtained by generating
additional MCMC\ samples.

For the BNP-IRT\ model, the standardized response residuals ranged from $%
-.21 $ to $.20$, meaning that the model had no outliers (i.e., all the
absolute standardized residuals were well below two). Globally, the model
fit analyses yielded criterion value $D(m)=2.76$ (with $\mathrm{Gof}(m)=.03$
and Penalty $P(m)=2.73$) for the $2,419$ responses in the data set. Also,
the BNP-IRT\ model attained an R-squared of one.

\begin{center}
---------------------

Insert Figure 1

---------------------
\end{center}

The estimated posterior means of the person$\,$ ability parameters were
found to be distributed with mean $.00,$ standard deviation $.46,$ minimum $%
-.66,$ and maximum $3.68$ for the $244$ persons. Figure 1 presents a box
plot of the marginal posterior distributions (full range, interquartile
range, median), for all the remaining parameters, including the
item-difficulty parameters and the slope coefficients of the covariates AGE,
FEMALE, and Miss:FEMALE. Parameter labels such as CERTIFICATE(1) and
CERTIFICATE(2) refer to the difficulty of the CERTIFICATE item, with respect
its rating categories 1 and 2, respectively. The most difficult item was
REMEDIAL(2) (with posterior median difficulty of $.27$), and the easiest
item was SECLANG(1) (posterior median difficulty $-1.81$). Also, the
covariates AGE and FEMALE were each found to have a significant positive
association with the rating response, since they had coefficients with 75\%
posterior intervals that excluded zero (this type of interpretation of
significance was justified by Li \&\ Lin, 2010\nocite{LiLin10}). The box
plot also presents the marginal posterior distributions for all the item and
covariate parameters in $\boldsymbol{\beta }_{\omega }$, the mixture
weights, and the variance parameters $\sigma _{\mu }$, $\sigma ^{2}$, and $%
\sigma _{\omega }^{2}$.

\section{Discussion\label{Discussion}}

In this chapter, we proposed and illustrated a practical and yet flexible
BNP-IRT model, which can provide robust estimates of person$\,$ ability and
item difficulty parameters. We demonstrated the suitability of the model
through the analysis of real polytomous item response data. The model showed
excellent predictive performance for the data, with no item response
outliers.

For the BNP-IRT model, a user-friendly and menu-driven software, entitled:\
"Bayesian Regression: Nonparametric and Parametric Models" is freely
downloadable from the authors website (Karabatsos, 2014a,b\nocite%
{Karabatsos14a}\nocite{Karabatsos14b}). The BNP-IRT model can be easily
specified by clicking the menu options "Specify New Model" and "Binary
infinite homoscedastic probits regression model." Afterwards, the response
variable, covariates, and prior parameters can be selected by the user.
Then, to run for data analysis, the user can click the "Run Posterior
Analysis" button to start the MCMC sampling algorithm in Section 4 for a
chosen number of iterations. Upon completion of the MCMC\ run, the software
automatically opens a text output file containing the results, which
includes summaries of the posterior distribution of the model obtained from
the MCMC samples. The software also allows the user to check for MCMC\
convergence through menu options that can be clicked to construct trace
plots or run a batch- mean analyses that produces 95\%\ Monte Carlo
confidence intervals of the posterior estimates of the model parameters.
Other menu options allow the user to construct plots (e.g., box plots) and
text with the estimated marginal\ posterior distributions of the model
parameters or residual plots and text reports the fit of the BNP-IRT model
in greater detail.

Currently, the software provides a choice of 59 statistical models,
including a large number of BNP\ regression models. The choice allows the
user to specify DP-mixture (or more generally, stick-breaking-mixture) IRT
models, with the mixing done either on the intercept parameter or the entire
vector of regression coefficient parameters.

An interesting extension of the BNP-IRT\ model would involve specifying the
kernel of the mixture by a cognitive model. For example, one may consider
the multinomial processing tree (MPT)\ model (e.g., Batchelder \&\ Riefer,
1999\nocite{BatchelderRiefer99}) with parameters that describe the latent
processes underlying the responses. Such an extension would provide a
flexible, infinite-mixture of cognitive models that allows cognitive
parameters to vary flexibly as a function of (infinitely-many)
covariate-dependent mixture weights.

\bigskip

\noindent {\Large References}

\begin{description}
\item Batchelder, W., \& Riefer, D. (1999). Theoretical and empirical review
of multinomial processing tree modeling. \textit{Psychonomic Bulletin and
Review}, \textit{6}, 57-86.

\item Begg, C., \& Gray, R. (1996). Calculation of polychotomous logistic
regression parameters using individualized regressions. \textit{Biometrika}, 
\textit{71}, 11-18.

\item DeIorio, M., M\"{u}ller, P., Rosner, G., \& MacEachern, S. (2004). An
ANOVA model for dependent random measures. \textit{Journal of the American
Statistical Association}, \textit{99}, 205-215.

\item Duncan, K., \& MacEachern, S. (2008). Nonparametric Bayesian modelling
for item response. \textit{Statistical Modelling}, \textit{8}, 41-66.

\item Farina, P., Quintana, F., Martin, E. S., \& Jara, A. (2009). \textit{A
dependent semiparametric Rasch model for the analysis of Chilean educational
data.} (Unpublished manuscript)

\item Ferguson, T. (1973). A Bayesian analysis of some nonparametric
problems. \textit{Annals of Statistics}, \textit{1}, 209-230.

\item Flegal, J., \& Jones, G. (2011). Implementing Markov chain Monte
Carlo: Estimating with confidence. In S. Brooks, A. Gelman, G. Jones, \& X.
Meng (Eds.), \textit{Handbook of Markov Chain Monte Carlo} (p. 175-197).
Boca Raton, FL: CRC.

\item Gelfand, A., \& Ghosh, J. (1998). Model choice: A minimum posterior
predictive loss approach. \textit{Biometrika}, \textit{85}, 1-11.

\item Geyer, C. (2011). Introduction to MCMC. In S. Brooks, A. Gelman, G.
Jones, \& X. Meng (Eds.), \textit{Handbook of Markov Chain Monte Carlo} (p.
3-48). Boca Raton, FL: CRC.

\item Ishwaran, H., \& James, L. (2001). Gibbs sampling methods for
stick-breaking priors. \textit{Journal of the American Statistical
Association}, \textit{96}, 161-173.

\item Kalli, M., Griffin, J., \& Walker, S. (2011). Slice sampling mixture
models. \textit{Statistics and Computing}, \textit{21}, 93-105.

\item Karabatsos, G. (2014a). Software for \textit{Bayesian Regression:
Nonparametric and parametric models}. http://www.uic.edu/
georgek/HomePage/BayesSoftware.html. University of Illinois-Chicago.

\item Karabatsos, G. (2014b). \textit{Bayesian Regression: Nonparametric and
parametric models. Software users manual.} http://www.uic.edu/
georgek/HomePage/BayesSoftware.html. University of Illinois-Chicago.

\item Karabatsos, G., \& Walker, S. (2012a). Adaptive-modal Bayesian
nonparametric regression. \textit{Electronic Journal of Statistics}, \textit{%
6}, 2038-2068.

\item Karabatsos, G., \& Walker, S. (2012b). Bayesian nonparametric mixed
random utility models. \textit{Computational Statistics and Data Analysis}, 
\textit{56(6)}, 1714-1722.

\item Laud, P., \& Ibrahim, J. (1995). Predictive model selection. \textit{%
Journal of the Royal Statistical Society}, \textit{Series B}, \textit{57},
247-262.

\item Li, Q., \& Lin, N. (2010). The Bayesian elastic net. \textit{Bayesian
Analysis}, \textit{5}, 151-170.

\item MacEachern, S. (1999). Dependent nonparametric processes. \textit{%
Proceedings of the Bayesian Statistical Sciences Section of the American
Statistical Association}, 50-55.

\item MacEachern, S. (2000). \textit{Dependent Dirichlet Processes} (Tech.
Rep.). The Ohio State University: Department of Statistics.

\item MacEachern, S. (2001). Decision theoretic aspects of dependent
nonparametric processes. In E. George (Ed.), \textit{Bayesian methods with
applications to science, policy and official statistics} (p. 551-560).
Creta: International Society for Bayesian Analysis.

\item Marron, J., \& Wand, M. (1992). Exact mean integrated squared error. 
\textit{The Annals of Statistics}, \textit{20}, 712-736.

\item Meyn, S., \& Tweedie, R. (1993). \textit{Markov chains and stochastic
stability}. London: Springer-Verlag.

\item Miyazaki, K., \& Hoshino, T. (2009). A Bayesian semiparametric item
response model with Dirichlet process priors. \textit{Psychometrika}, 
\textit{74}, 375-393.

\item Mukhopadhyay, S., \& Gelfand, A. (1997). Dirichlet process mixed
generalized linear models. \textit{Journal of the American Statistical
Association}, \textit{92}, 633-639.

\item National Research Council. (2010). \textit{Preparing teachers:
Building evidence for sound policy.} Washington, DC: National Academies
Press.

\item Neal, R. (2003). Slice sampling (with discussion). \textit{Annals of
Statistics}, \textit{31}, 705-767.

\item Nummelin, E. (1984). \textit{General irreducible Markov chains and
non-negative operators.} London: Cambridge University Press.

\item Qin, L. (1998). \textit{Nonparametric Bayesian models for item
response data} (Ph.D. Thesis). Unpublished doctoral dissertation, The Ohio
State University.

\item Rasch, G. (1960). \textit{Probabilistic models for some intelligence
and achievement tests. }Copenhagen: Danish Institute for Educational
Research: Danish Institute for Educational Research (Expanded edition, 1980.
Chicago: University of Chicago Press).

\item Robert, C., \& Casella, G. (2004). \textit{Monte Carlo statistical
methods (second edition).} New York: Springer.

\item Roberts, G., \& Rosenthal, J. (2004). General state space Markov
chains and MCMC algorithms. \textit{Probability Surveys}, \textit{1}, 20-71.

\item Rost, J. (1990). Rasch models in latent classes: An integration of two
approaches to item analysis. \textit{Applied Psychological Measurement}, 
\textit{14}, 271-282.

\item Rost, J. (1991). A logistic mixture distribution model for
polychotomous item responses. \textit{British Journal of Mathematical and
Statistical Psychology},\textit{\ 44}, 75-92.

\item San Mart\'{\i}n, E., Jara, A., Rolin, J.-M., \& Mouchart, M. (2011).
On the Bayesian nonparametric generalization of IRT-type models. \textit{%
Psychometrika}, \textit{76}, 385-409.

\item San Mart\'{\i}n, E., Rolin, J.-M., \& Castro, L. M. (2013).
Identification of the 1PL model with guessing parameter: Parametric and
semi-parametric results. \textit{Psychometrika}, \textit{78}, 341-379.

\item Sethuraman, J. (1994). A constructive definition of Dirichlet priors. 
\textit{Statistica Sinica}, \textit{4}, 639-650.
\end{description}

\newpage

\begin{center}
\textbf{Figure Caption}
\end{center}

\textbf{Figure 1.} For the BNP-IRT\ model, a box plot of the marginal
posterior distributions of the item, covariate, and prior parameters. For
each of these model parameters, the box plot presents the range,
interquartile range, and median.

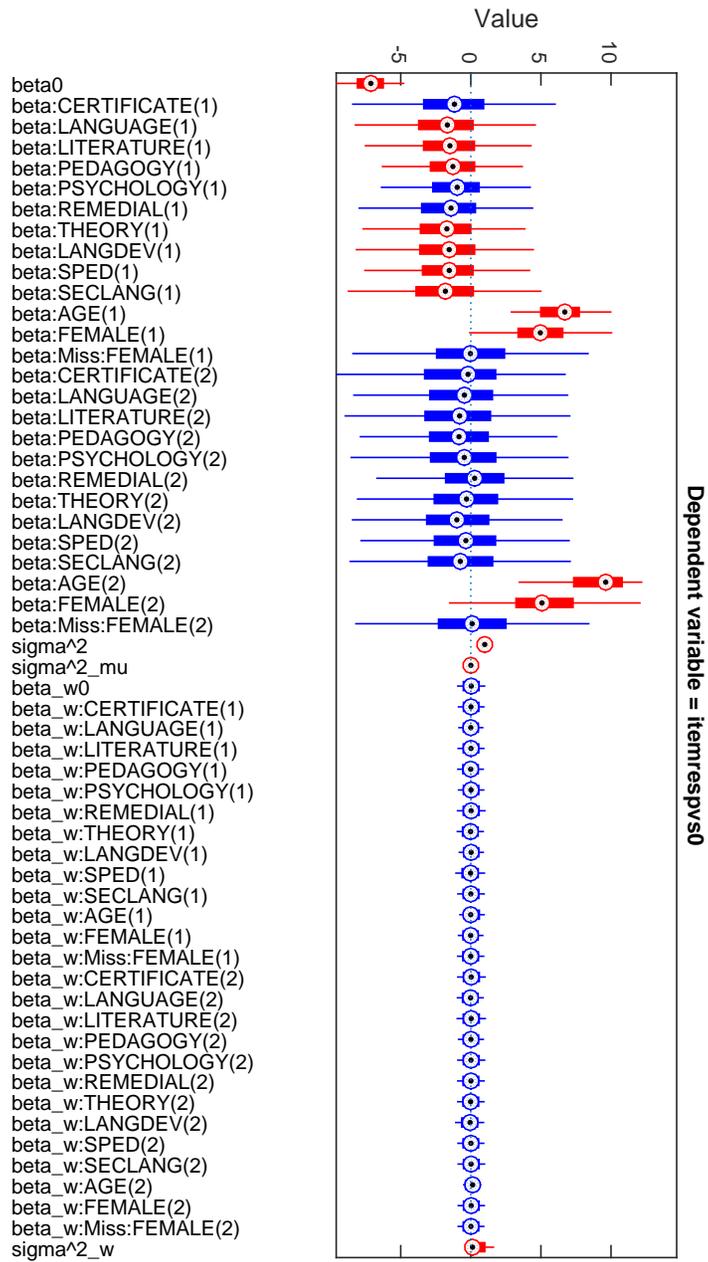
\end{document}